\documentclass[twocolumn, showpacs, prl,unsortedaddress]{revtex4}

\usepackage{graphicx}

\usepackage{dcolumn}

\usepackage{color}
\usepackage{bm}

\begin{document}

\title{Dependence of Effective Mass on Spin and Valley Degrees of Freedom}

\date{\today}

\author{T.\ Gokmen}

\author{Medini\ Padmanabhan}

\author{M.\ Shayegan}

\affiliation{Department of Electrical Engineering, Princeton
University, Princeton, NJ 08544}

\begin{abstract}

We measure the effective mass ($m^{*}$) of interacting
two-dimensional electrons confined to an AlAs quantum well while we
change the conduction-band valley occupation and the spin
polarization via the application of strain and magnetic field,
respectively. Compared to its band value, $m^{*}$ is enhanced unless
the electrons are fully valley $and$ spin polarized. Incidentally,
in the fully spin- and valley-polarized regime, the electron system
exhibits an insulating behavior.
\end{abstract}


\maketitle

In Fermi liquid theory, interacting electrons can be treated as
non-interacting quasi-particles with a re-normalized effective mass
($m^{*}$). The parameter $m^{*}$ has been studied extensively
\cite{SmithPRL72, PanPRB99, ShashkinPRL03, TanPRL2005,
PadmanabhanPRL08} for various two-dimensional electron systems
(2DESs) as a function of interaction strength, which is
characterized by the ratio $r_{s}$ of the Coulomb energy to Fermi
energy. In the highly interacting, dilute regime ($r_{s}>3$),
$m^{*}$ is typically significantly enhanced compared to its band
value ($m_{b}$) and tends to increase with increasing $r_{s}$
\cite{SmithPRL72, PanPRB99, ShashkinPRL03, TanPRL2005,
PadmanabhanPRL08, KwonPRB94, AsgariPRB05, AsgariPRB2006,
GangadharaiahPRL05, ZhangPRL05}. A question of particular interest
is the dependence of $m^{*}$ on the 2D electrons' spin and valley
degrees of freedom as these affect the exchange interaction.
Measurements on 2DESs in Si-MOSFETs (metal-oxide-semiconductor field
effect transistors), have indicated that the $m^{*}$ enhancement is
independent of the degree of spin-polarization \cite{ShashkinPRL03}.
On the other hand, recent $m^{*}$ measurements for 2D electrons
confined to highly strained AlAs quantum wells revealed that, when
the 2DES is fully spin-polarized, $m^{*}$ is $suppressed$ down to
values near or even slightly below $m_b$ \cite{PadmanabhanPRL08}.
The reason for this apparent discrepancy might be that the 2D
electrons in the Si-MOSFET case occupy two conduction-band valleys
\cite{GangadharaiahPRL05} while the 2DES studied in Ref.
\cite{PadmanabhanPRL08} is a single-valley system. Here we
experimentally study $m^{*}$ in an AlAs 2DES while we tune the
valley and spin degrees of freedom in a single sample. The results,
summarized in Fig. 1, provide direct and conclusive evidence that
the valley and spin degrees of freedom are indeed $both$ important
in $m^{*}$ re-normalization. If the electrons are only partially
valley and/or spin polarized, then their $m^{*}$ is enhanced with
respect to $m_b$. But if they are fully valley $and$ spin polarized,
then $m^{*}$ is suppressed. The fully spin and valley polarized
regime is incidentally also the regime where the 2DES shows an
insulating behavior.

\begin{figure}
\centering
\includegraphics[scale=0.88]{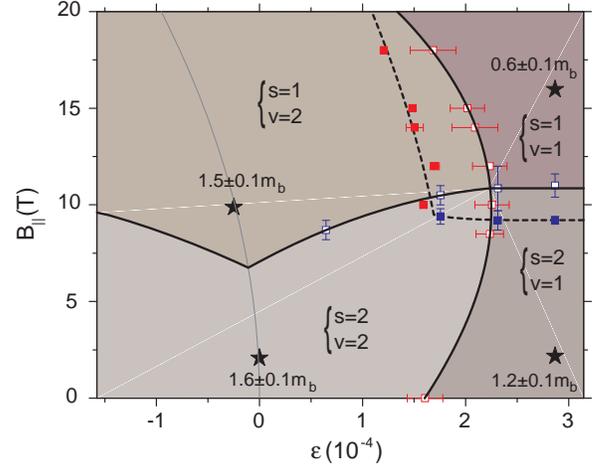}
\caption{(Color online) The spin (s) and valley (v) occupations for
density $n=3.8$$\times$$10^{11}$ cm$^{-2}$ are shown as a function
of $B_{\|}$ and strain; the thick, solid lines mark the boundaries
and the indices 1 and 2 indicate, respectively, whether one or two
spin (and/or valley) subbands are occupied. Dashed lines mark the
boundary of the upper-right region where the sample exhibits an
insulating behavior. Stars mark the ($B_{\|}$, $\epsilon$) values at
which the indicated effective mass was measured.}
\end{figure}

\begin{figure}
\centering
\includegraphics[scale=0.87]{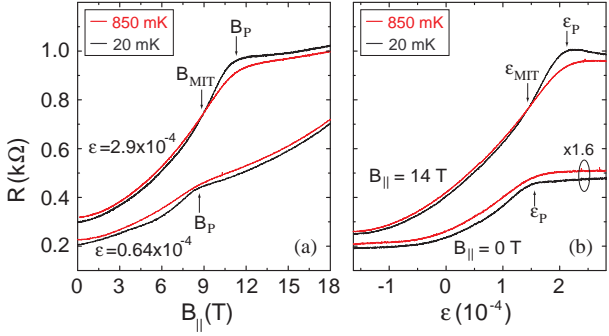}
\caption{(Color online) Sample resistance is shown as a function of
(a) $B_\|$ and (b) $\epsilon$ at two temperatures. For each pair of
traces in (a), $\epsilon$ is kept constant while for each pair in
(b) $B_\|$ is kept fixed at the indicated values.}
\end{figure}

We studied a high-mobility 2DES confined to an 11 nm-thick,
modulation-doped AlAs layer, grown by molecular beam epitaxy on a
(001) GaAs substrate \cite{ShayeganPSS2006}. In this sample, the 2D
electrons occupy two energetically degenerate conduction-band
valleys with elliptical Fermi contours \cite{ShayeganPSS2006}, each
centered at an X point of the Brillouin zone, and with an
anisotropic mass (longitudinal mass $m_{l}=1.05$ and transverse mass
$m_{t}=0.20$, in units of free electron mass, $m_{e}$). The two
valleys have their major axes either along the [100] or the [010]
crystal directions; we denote them as [100] and [010] valleys,
respectively. Their degeneracy can be lifted by applying a symmetry
breaking strain $\epsilon =\epsilon _{[100]} - \epsilon _{[010]}$,
where $\epsilon_{[100]}$ and $\epsilon_{[010]}$ are the strain
values along the [100] and [010] directions. Electrons are
transferred from the [100] valley to the [010] valley for positive
$\epsilon$ and vice versa for negative $\epsilon$
\cite{ShayeganPSS2006}. We control the valley occupation using
$\epsilon$ and the spin occupation (polarization) via the
application of magnetic field. We studied several samples; here we
focus on data from one sample patterned with a
60$\times$60$\mu$m$^2$ square mesa whose edges are aligned along the
[100] and [010] directions, and the current is passed along [010].
The sample is glued to the side of a stacked piezo actuator allowing
us to apply in-plane strain which can be controlled and measured
\textit{in situ} \cite{ShayeganPSS2006}. The magneto-resistance
measurements were performed in a dilution refrigerator with a base
temperature ($T$) of 0.02 K and equipped with a tilting stage,
allowing the angle $\theta$ between the sample normal and the
magnetic field to be varied \textit{in situ}.

\begin{figure}
\centering
\includegraphics[scale=0.76]{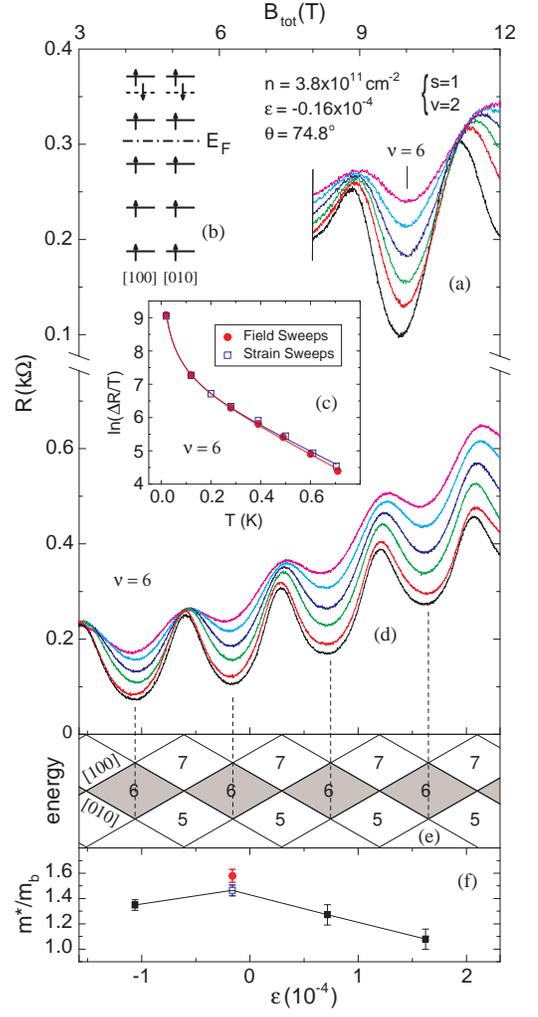}
\caption{(Color online) Summary of measurements for the (s=1, v=2)
case. (a) and (d): Field and strain traces taken at different
temperatures ranging from 0.02K to 0.7K (bottom to top). (b): Energy
level diagram. (c): Dingle fits at $\nu=6$. (e): Landau level fan
diagram as a function of $\epsilon$. (f): Deduced values of $m^{*}$
as a function of $\epsilon$.}
\end{figure}

\begin{figure*}
\centering
\includegraphics[scale=0.56]{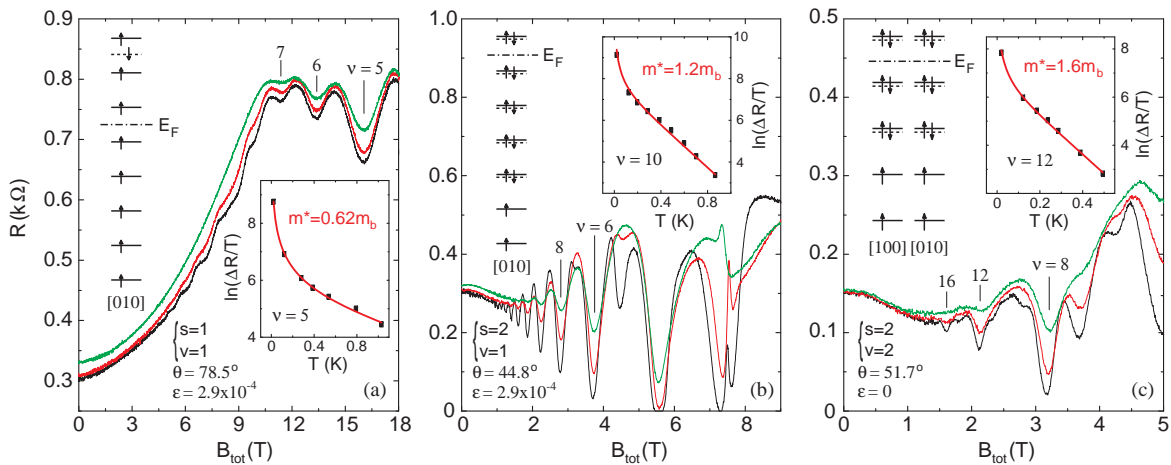}
\caption{(Color online) Magneto-resistance traces, Dingle fits, and
energy level diagrams for (s=1, v=1), (s=2, v=1) and (s=2, v=2). The
black, red and green traces were taken at (a) 0.02, 0.39, 1.03; (b)
0.02, 0.60, 0.87; (c) 0.02, 0.24, 0.39K.}
\end{figure*}

The $B_{\|}$ vs. $\epsilon$ phase diagram of Fig. 1 highlights the
important features of our results ($B_{\|}$ is the magnetic field
applied parallel to the 2D plane). In this figure, the thick,
solid lines mark the boundaries between the four possible spin and
valley subband occupations for our sample, which had a density
$n=3.8\times10^{11}$ cm$^{-2}$; we denote the occupations by (s=2,
v=2), (s=2, v=1), (s=1, v=2) and (s=1, v=1), where s and v stand
for spin and valley, and 1 and 2 denote the number of spin/valley
subbands that are occupied \cite{Footnote1}. We determined these
boundaries from measurements of sample resistance ($R$) vs. either
$B_{\|}$ or $\epsilon$ while the other parameter is kept fixed.
Examples of such data are shown in Fig. 2 at two temperatures. In
each set of traces, $R$ increases and shows a kink at a field or
strain, $B_{P}$ or $\epsilon_{P}$, which marks the onset of full
spin or valley polarization, respectively
\cite{GunawanNature2007}. The $B_{P}$ and $\epsilon_{P}$ values
are shown in Fig. 2 by open square symbols. Note in Fig. 1 that
the boundaries are not simple horizontal or vertical lines. This
is because the Fermi energy ($E_{F}$) depends on the size of the
(partial) occupation of the valleys or spins. Moreover, in a 2DES
with finite layer thickness there is a shift, with $B_{\|}$, of
$\epsilon$ at which the valleys are equally occupied, or are
"balanced", as indicated by the thin grey line going through
$\epsilon=0$ in Fig. 1. In our experiments, where $B_{\|}$ is
along [100], the balanced point shifts to negative values of
$\epsilon$ at high $B_{\|}$ because, $B_{\|}$ shifts the energy of
the [100] valley above the [010] valley \cite{Gokmen2008}. We
emphasize that, although we can measure and explain the boundaries
shown in Fig. 1 accurately, their precise locations are not
important for the conclusions of our manuscript. As we describe
below, we carefully choose the conditions of our sample ($B_{\|}$,
$\epsilon$, and $\theta$) when we perform our $m^{*}$ measurements
in different quadrants of Fig. 1. The stars show the coordinates
where we measured $m^{*}$ in these four quadrants, and the deduced
$m^{*}$ values are indicated next to the stars. Finally, the
closed squares and the dashed lines in Fig. 1 mark the boundary
for the apparent metal-insulator transition observed at this
density in our sample (see Fig. 2 data). The 2DES shows an
insulating behavior only as it becomes sufficiently spin and
valley polarized \cite{GunawanNature2007}.

In the remainder of the paper we describe our determination of
$m^{*}$, from the $T$-dependence of the Shubnikov-de Haas (SdH)
oscillations, in different quadrants of Fig. 1. Figure 3 shows the
data for the (s=1, v=2) case which were taken at
$\theta=74.8^{\circ}$ and $\epsilon=-0.16\times10^{-4}$. These
values for $\theta$ and $\epsilon$ were chosen carefully so that the
lowest eight Landau levels (LLs) are spin polarized and the two
valleys have equal densities at the LL filling factor $\nu=6$, as
shown in Fig. 3(b). Figure 3(a) shows the SdH oscillation near
$\nu=6$ at several T. To deduce $m^{*}$, we analyzed the
$T$-dependence of the amplitude of SdH oscillations using the
standard Dingle expression \cite{DinglePRSL52}: $\Delta R/R_{o} = 8
exp(-\pi/\omega_{c}\tau_{q})\xi/sinh(\xi)$, where the factor
$\xi/sinh(\xi)$ represents the \textit{T}-induced damping,
$\xi=2\pi^{2}k_{B}T/\hbar\omega_{c}$, and
\textit{$\omega_{c}=eB_{\bot}/m^{*}$} is the cyclotron frequency,
$R_{o}$ is the non-oscillatory component of the resistance, and
$\tau_{q}$ is the single-particle (quantum) lifetime. Here we make
the common assumption that both $R_{o}$ and $\tau_{q}$ are
$T$-independent, and will return to its justification later in the
Letter. Figure 3(c) shows that the data points (red circles) fit the
Dingle expression (red line) quite well, yielding $m^{*}=1.58m_{b}$,
where $m_{b}=\sqrt{m_{l}m_{t}}=0.46m_e$.

In Fig. 3(d) we show an alternative set of data from which we
determine $m^{*}$ for the (s=1, v=2) case. In contrast to the
standard SdH oscillations shown in Fig. 3(a), where we vary the
magnetic field at constant $\epsilon$, in Fig. 3(d) we fix the value
of field so that the 2DES remains at $\nu=6$ and continuously change
$\epsilon$ (see Ref. \cite{GunawanPRL2006} for details). With
applied $\epsilon$, the LLs for the [100] and [010] valleys cross
each other, as the fan diagram in Fig. 3(e) indicates, and $R$
exhibits minima as $E_{F}$ goes through consecutive energy gaps, and
maxima as it coincides with the LL crossings. We analyzed the
$T$-dependence of the amplitudes of these resistance oscillations to
deduce $m^{*}$. In Fig. 3(c) we show a plot of $\Delta R/T$ vs. $T$
(blue squares) for the oscillation centered around
$\epsilon=-0.16\times10^{-4}$; the data fit the Dingle expression
quite well (blue line), and yield $m^{*}=1.46m_{b}$ which is close
to the value obtained from the field-sweep data of Fig. 3(a). In a
similar fashion, we analyzed the $T$-dependence of the other
resistance oscillations observed in Fig. 3(d); the data fit the
Dingle expression well, and provide the $m^{*}$ values plotted in
Fig. 3(f). Evidently, $m^{*}$ is smaller when we "valley polarize"
the 2DES: note that the valley polarization ($P_{V}$), defined as
the difference between the [010] and [100] valley populations
divided by the total 2DES density, is equal to -0.33, 0, 0.33, and
0.67 for the four data points shown in Fig. 3(f) (from left to
right).

The data for the (s=1, v=1) case are shown in Fig. 4(a). Here
$\epsilon$ is large enough so that only the [010] valley is
occupied. Moreover, by tilting the sample to a sufficiently large
$\theta$ we ensure that the lowest seven LLs are spin polarized, as
indicated in Fig. 4(a) energy level diagram. Note the qualitative
similarity of the field traces in Figs. 4(a) and 2(b): $R$ shows a
pronounced increase with field and then saturates above $\sim 11$T
once the 2DES is completely spin polarized. Since the sample is not
completely parallel to the field in Fig. 4(a), however, there are
SdH oscillations which become well resolved after the full spin
polarization; these are the oscillations from which we deduce
$m^{*}$. The inset to Fig. 4(a) shows the $T$-dependence of the SdH
oscillation centered around $\nu=5$. From the fit to the Dingle
expression we obtain $m^{*}$=0.62$m_b$ at this $\nu=5$; data at
$\nu=6$ yield $m^{*}$=0.60$m_b$.

Figure 4(b) shows data for the (s=2, v=1) case. Here $\epsilon$ is
chosen large enough so that the system is completely valley
polarized and $\theta$ is adjusted such that the LLs at odd $\nu$
are at coincidence. The corresponding LL energy diagram is shown in
Fig. 4(b) inset. Consistent with this diagram, in Fig. 4(b) traces,
resistance minima at even $\nu$ are strong while the minima at odd
$\nu$ are either weak or entirely absent (at high $\nu$), or are
accompanied by spikes (e.g., at $\nu=3$)
\cite{DePoortereScience2000}. By fitting the amplitude of the SdH
oscillations near $\nu=$ 8, 10, and 12 to the Dingle expression, we
obtain $m^{*}$=1.16, 1.22, and 1.22$m_{b}$, respectively (see Fig.
4(b) inset as an example). This observation implies that $m^{*}$
does not depend on the spin polarization ($P_{S}$), defined as the
difference between the spin up and down populations divided by the
total electron density; such independence is not surprising since in
the range of $8 \leq \nu \leq 12$, $P_{S}$ changes only from 0.17 to
0.25.

In Fig. 4(c), we show data for (s=2, v=2). Here $\epsilon=0$ and
$\theta$ is chosen so that the spin-up and spin-down LLs are at
coincidence, as illustrated in the energy diagram in Fig. 4(c). The
strongest resistance minima are observed at $\nu=8$, 12, and 16. We
obtain $m^{*}=1.58m_{b}$ from the fit of SdH oscillation amplitude
around $\nu=12$ to the Dingle expression, as shown in Fig. 4(c)
inset.

In our determination of $m^*$ from the SdH oscillations, we have
assumed that $R_{o}$ and $\tau_{q}$ are $T$ independent. It is
apparent in Figs. 3-4, however, that $R_{o}$ has some dependence on
$T$. To take this into account, we also analyzed our data by
defining $R_{o}$ as the average resistance value near a SdH
oscillation, and assumed that the relative $T$-dependence of
$\tau_{q}$ is similar to but has half the size of the relative
$T$-dependence of $R_{o}$ \cite{AdamovPRB2006}. The deduced $m^*$
from such analysis are about 10\% smaller than the values reported
above, indicating that the main conclusions of our study remain
intact.

Our results presented in Figs. 3 and 4, and summarized in Fig. 1,
provide direct experimental evidence for the dependence of $m^{*}$
on valley and spin polarization of the 2DES. This is intuitively
plausible since $m^{*}$ re-normalization in an interacting 2DES
partly stems from the exchange interaction which depends on the spin
and valley degrees of freedom. Several features of our data are
noteworthy. First, we observe the largest $m^{*}$ enhancement for
the (s=2, v=2) case. As we fully spin polarize the 2DES while
keeping v=2, we observe little dependence of $m^{*}$ on
spin-polarization. This observation is consistent with the
experimental results of Shashkin \textit{et al}.
\cite{ShashkinPRL03} who reported that $m^{*}$ does not depend on
the degree of spin-polarization in Si-MOSFET 2DESs where the
electrons occupy two valleys. Theoretical justification for this
independence is provided in Ref. \cite{GangadharaiahPRL05} where it
is argued that in a two-valley 2DES the dependence of $m^{*}$ on the
spin polarization is likely to be too weak to be experimentally
measurable. On the other hand, when we keep s=2 and fully valley
polarize the 2DES, we observe about 20\% reduction in $m^{*}$. The
reason for this reduction is not clear, but we note that the spin
and valley degrees of freedom are not identical. Second, data of
Fig. 3(f) suggest that, if s=1, then there is reduction of $m^{*}$
with increasing valley polarization \cite{Footnote2}. If one treats
the spin and valley degrees of freedom as qualitatively similar,
this observation is consistent with the theoretical work of Ref.
\cite{ZhangPRL05} where, for the majority spin electrons, a
reduction of $m^{*}$ with increasing spin polarization has been
reported in a single-valley 2DES. Third, we find that $m^{*}$ is
significantly reduced, to values even below $m_b$ when the 2DES is
fully spin and valley polarized \cite{PadmanabhanPRL08}, implying
that the exchange interaction is notably different when the spin and
valley degrees of freedom are $both$ frozen out.

Finally, our data imply that there may be a link between the $m^{*}$
suppression and the metal-insulator (MIT) transition observed in our
2DES. As seen in Fig. 1, the strong $m^{*}$ suppression and the
insulating phase both occur in the regime where the 2DES is spin and
valley polarized. It is important to note, however, that our $m^{*}$
data were all taken in the presence of a perpendicular component of
the magnetic field. Such a field could suppress the insulating
behavior; this can be seen, e.g., in Fig. 4(a) where the MIT
observed around $B_{\|}$ = 9T in the upper traces of Fig. 2(a) is no
longer visible.

In summary, measurements in an AlAs 2DES reveal variations of
$m^{*}$ with spin and valley polarizations. Some of the variations
we observe can be qualitatively explained by the existing
calculations \cite{GangadharaiahPRL05, ZhangPRL05}. A quantitative
understanding of our observations, including the apparent link
between the $m^{*}$ suppression and the insulating phase in the
fully spin and valley polarized regime, awaits future work.

We thank the NSF for support, D.L. Maslov, S. Das Sarma, and E.
Tutuc for helpful discussions. Part of this work was done at the
NHMFL, Tallahassee, which is also supported by the NSF. We thank E.
Palm, T. Murphy, J. Jaroszynski, S. Hannahs and G. Jones for
assistance.

\break


\begin{references}
\bibitem{Footnote1}
For the (s=2, v=1) case, $r_s$ = 8.0 for our sample.
\bibitem{Footnote2}
Analogously, one might expect that, when v=1, $m^{*}$ should
decrease with increasing spin-polarization. We cannot rule out such
possibility since our data is experimentally limited to only a small
range of spin polarization 0.17 $<$ $P_{S}$ $<$ 0.25.
\end{references}
\end{document}